\documentclass[conference]{IEEEtran}

\usepackage{cite}
\usepackage{amsmath,amssymb,amsfonts}
\usepackage{algorithmic}
\usepackage{graphicx}
\usepackage{textcomp}
\usepackage{xcolor}
\usepackage{booktabs}
\usepackage{listings}
\lstdefinestyle{javacode}{
    basicstyle=\ttfamily\scriptsize,
    keywordstyle=\color{blue!70!black}\bfseries,
    commentstyle=\color{green!50!black},
    stringstyle=\color{red!60!black},
    showstringspaces=false,
    breaklines=false,
    columns=fullflexible,
    keepspaces=true,
    tabsize=2,
    aboveskip=2pt,
    belowskip=2pt,
}
\usepackage[hidelinks]{hyperref}
\def\BibTeX{{\rm B\kern-.05em{\sc i\kern-.025em b}\kern-.08em
    T\kern-.1667em\lower.7ex\hbox{E}\kern-.125emX}}
\begin{document}

\title{Fail-Aware and Explainable Test Oracle Prediction}

\author{
\IEEEauthorblockN{
Yue Zhao,
Binish Tanveer,
Jelena Zdravkovic
}

\IEEEauthorblockA{
Department of Computer and Systems Sciences,
Stockholm University, Stockholm, Sweden\\
\{yue.zhao, binish.tanveer, jelenaz\}@dsv.su.se
}
}
\maketitle

\begin{abstract}

Despite their central role in fault detection, test oracles remain challenging to construct effectively. Recent learning-based methods address this challenge by automatically generating test assertions, yet even if syntactically correct, they are often ineffective in revealing bugs. Rather than generating assertions, this study explores a different approach by training a model to directly predict whether a given test prefix passes or fails. 

We present FOCAL, an emerging code LLM-based discriminative oracle predictor. It learns from labeled pairs of test prefixes and methods under test, employs losses that emphasize failing cases during training, and grounds its predictions in statement-level behavioral evidence. Compared with the baseline method SEER, we substantially improve performance on failing cases for unseen projects and provide richer explanations.

A preliminary evaluation on fault-detection benchmarks and automated test-generation artifacts shows that our approach is highly accurate within its training distribution and substantially improves failure detection on previously unseen projects where prior discriminative oracles collapse. Moreover, the highlighted statements are supported by behavioral explanation checks. These early results suggest that fail-aware discriminative oracle prediction can complement existing approaches such as fuzzing, search-based testing, and LLM-based test generation. These techniques produce test prefixes at scale but often lack fault-oriented oracles. In future work, FOCAL could take generated test prefixes and attach fault-aware predicted oracles to them, turning high-volume input generation into executable tests that are more likely to expose semantic failures.

\end{abstract}


\section{Introduction}

As a cornerstone of modern software quality assurance, unit testing is widely used to check the correctness of individual program units and provide fast, localized feedback during development~\cite{myers2004art, ammann2017introduction}. A unit test typically combines a \emph{test prefix} with a \emph{test oracle}. The prefix executes the setup and stimulus code for the Method Under Test (MUT), and the test oracle determines whether the observed behavior conforms to the required specification~\cite{dinella2022toga}. Automated test prefix generation has become practical through tools such as EvoSuite~\cite{fraser2013evosuite} and Randoop \cite{pacheco2007randoop}. In contrast, test oracle construction remains one of the most persistent and costly challenges, widely recognized as the \textit{oracle problem} \cite{barr2014oracle,zhang2015assertions}.

Automated unit testing constructs oracles through different strategies. Fuzzing tools~\cite{sutton2007fuzzing} use runtime signals such as crashes, hangs, and sanitizer violations as implicit oracles to capture visible runtime failures. Search-based tools record observed behavior as regression assertions, which detect future behavioral drift and take current behavior as the reference~\cite{shamshiri2015automatically}. Neural-based assertion generation tools~\cite{watson2020learning,yu2022automated} train models to synthesize \texttt{assert} statements that emphasize textual similarity to developer-written assertions. LLM-based and hybrid approaches~\cite{lemieux2023codamosa} produce readable oracles following common patterns. More precise judgments can be obtained when specifications, contracts, or reference implementations are available and reliable~\cite{blasi2018translating}, but such conditions are uncommon in many real-world projects. These approaches provide useful oracle signals, but they still leave a key gap in our target setting. Given a test prefix without an oracle and an MUT, there is often no explicit judgment of whether the observed behavior reveals a fault.

SEER~\cite{ibrahimzada2022perfect}, a new oracle construction effort, recasts oracle construction as an explicit judgment. Given a test prefix and an MUT, predict whether the pair should \textsc{Pass} or \textsc{Fail}. The verdict itself plays the role of the oracle, without writing an assertion into the test file. This formulation directly targets the gap above. However, our evaluation of SEER~\cite{ibrahimzada2022perfect} under an unseen project exposed an important decoupling. SEER achieves 86.72\% overall accuracy, but this aggregate score does not translate into effective failure detection, with a \textsc{Fail} recall of only 2.95\% and a \textsc{Fail} F1 of 3.87\%.

To our knowledge, previous evaluations of discriminative oracles have not clearly examined whether high overall accuracy also translates into effective failure detection. This finding is important because high overall scores on pass-dominant data can coexist with very weak sensitivity to the cases most directly tied to fault detection. We therefore argue for a fail-aware research direction for learned oracles. Failure detection should be optimized and reported explicitly as a primary evaluation target, rather than inferred from aggregate performance alone. Our contributions are as follows:

\begin{itemize}
\item We revisit learned test oracle prediction from a fail-aware perspective and show that aggregate accuracy tends to hide weak detection of failing cases in unseen-project settings. This motivates reporting fail-aware metrics as a primary evaluation target for learned test oracles.
\item We present FOCAL, a fail-aware discriminative oracle predictor that encodes the test prefix and MUT separately with a shared CodeT5~\cite{wang2021codet5} encoder adapted via Low-Rank Adaptation (LoRA). FOCAL combines relation-aware pair representation with focal classification and margin ranking to improve sensitivity to failing cases.
\item We provide statement-level explanations for FOCAL's \textsc{Fail} predictions using Integrated Gradients (IG)~\cite{sundararajan2017axiomatic} and evaluate the selected evidence with deletion, keep-only, and counterfactual checks. These checks offer a behavioral basis for inspecting FOCAL judgments.
\end{itemize}

\section{Related Work and Positioning}\label{sec:related}

Research on test oracles spans a broad range of mechanisms, from hand-crafted rules and runtime signals to learned models that generate or judge program behavior. These mechanisms can be grouped into three categories: \emph{traditional oracle mechanisms}, \emph{learning-based test and assertion generation}, and \emph{discriminative oracle prediction}.

\subsection{Traditional Oracle Strategies}

The earliest and still most widely used oracles rely on external references, recorded behavior, or runtime symptoms. 

\textbf{Specified or derived oracles:} Oracles drawn from formal contracts, natural-language specifications~\cite{motwani2023better}, differential testing against alternative implementations~\cite{nilizadeh2021exploring}, or trusted reference implementations~\cite{dinella2022toga} can deliver precise verdicts when applicable. Their main limitation is availability, such resources are frequently absent in real-world projects and in cold-start testing scenarios.

\textbf{Regression oracles:} Search-based tools construct inputs or prefixes that improve structural coverage and record observed behavior as regression assertions. The resulting suites detect later behavioral drift but cannot reveal faults already present in the version under measurement, since current behavior is taken as the oracle~\cite{almasi2017industrial}.

\textbf{Implicit oracles:} Fuzzing tools such as AFL++~\cite{fioraldi2020afl++}, libFuzzer~\cite{serebryany2016continuous}, and JQF~\cite{padhye2019jqf} drive programs with large input volumes and report failures when executions crash, hang, or trigger sanitizers. This oracle requires no specification and scales to long campaigns, but it observes mainly failures that surface at runtime, whereas silent semantic faults can remain invisible.

\subsection{Learning-Based Test and Assertion Generation}
A second category of work uses learning-based models to produce oracle artifacts, typically in the form of executable assertion text accompanying a test prefix.

\textbf{Neural assertion synthesis:} Sequence-to-sequence models such as ATLAS~\cite{watson2020learning} and AthenaTest~\cite{tufano2022generating} are trained to synthesize \texttt{assert} statements directly from a test prefix. The output is executable, but the training objective rewards reproducing assertion syntax and frequent patterns rather than deciding whether the underlying behavior is semantically correct.

\textbf{LLM-based and hybrid test generation:} More recent work, such as A3Test~\cite{alagarsamy2024a3test} and TOGLL~\cite{hossain2025togll}, fine-tunes large language models (LLMs) for assertion synthesis, and other approaches combine LLMs with search-based testing to produce readable tests with syntactically valid assertions~\cite{schafer2023empirical}. However, empirical studies report that generated assertions concentrate on a small number of recurring patterns and reproduce observed behavior more often than they expose faults~\cite{siddiq2024using}. Hybrid configurations that combine LLMs with search-based software testing improve fluency and coverage, yet do not by themselves yield a fault-revealing oracle.

\subsection{Discriminative Oracle Prediction}
Discriminative oracle prediction addresses the gap left by the previous two categories. It applies when no specification, reference implementation, or prior version is available, where traditional mechanisms often rely on implicit runtime signals that can miss silent faults. It also avoids relying on generated assertions, which previous empirical work shows may reproduce observed behavior rather than expose faults~\cite{siddiq2024using}. By reconstructing the oracle to a single \textsc{Pass}/\textsc{Fail} judgment, this formulation turns fault detection into an explicit prediction target, rather than an outcome mediated by assertion quality or runtime symptoms.

\textbf{SEER:} SEER is a representative early approach that instantiates this formulation~\cite{ibrahimzada2022perfect}. It demonstrates the feasibility of learning a verdict directly, but exhibits two limitations that constrain practical use: weak performance on failing cases, where fault detection matters most, and the absence of trustworthy explanations for its predictions.

\textbf{FOCAL:} We present FOCAL, a fail-aware discriminative oracle predictor organized around three dimensions: \emph{representation}, \emph{optimization}, and \emph{explainability}. For \emph{representation}, FOCAL encodes the test prefix and MUT separately using a shared CodeT5 encoder adapted with LoRA. For \emph{optimization}, it combines focal classification~\cite{lin2017focal} with margin ranking to increase sensitivity to failing cases. For \emph{explainability}, it derives statement-level evidence using IG and validates the evidence through behavioral checks, including counterfactual edits~\cite{wachter2017counterfactual}.

We do not suggest that discriminative oracles replace generative or human-written oracles. Instead, we contend that they provide a complementary alternative, particularly in scenarios where a fault-detection oracle is needed but executable assertions are unavailable or impractical to write. FOCAL targets testing settings where executable test prefixes can be produced at scale, but reliable fault-detection oracles are missing or costly to obtain. 

\section{FOCAL Overview}

\begin{figure*}[t]
    \centering
    \includegraphics[width=0.85\linewidth]{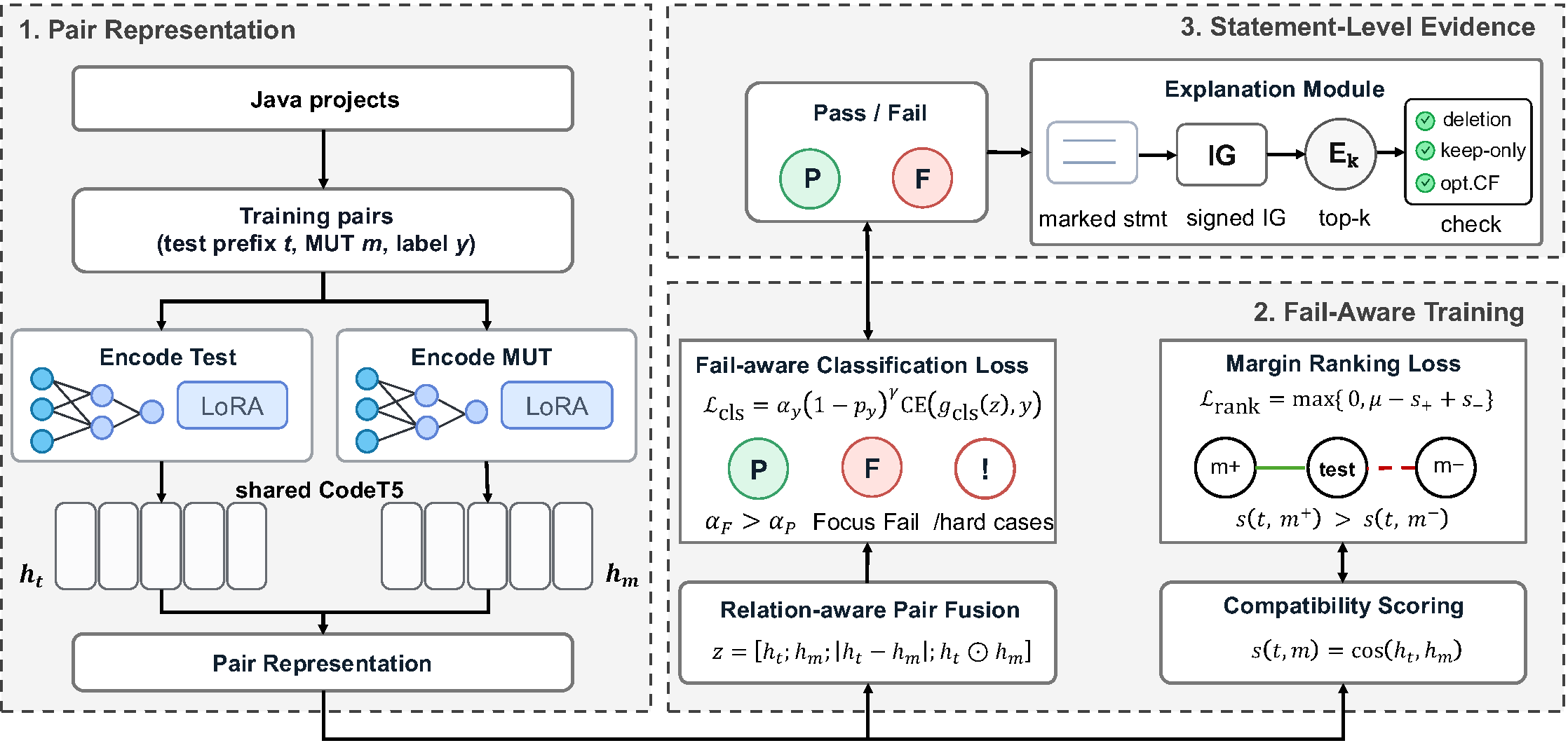}
    \caption{Overview of FOCAL}
    \label{fig:overview}
\end{figure*}

FOCAL is a fail-aware discriminative oracle predictor that learns a classifier $f(t,m) \rightarrow \{\textsc{Pass}, \textsc{Fail}\}$ for a given test prefix $t$ and MUT $m$.
As shown in Figure~\ref{fig:overview}, FOCAL first encodes the test prefix and the MUT separately with a shared CodeT5 encoder adapted by LoRA. It then derives a relation-aware pair representation for \textsc{Pass}/\textsc{Fail} prediction and a compatibility score for margin ranking supervision. For explanation, FOCAL analyzes pairs predicted as \textsc{Fail}, maps IG attributions to test and MUT statements, and validates the selected evidence with deletion, keep-only, and lightweight counterfactual checks.

\subsection{Pair Representation}
FOCAL uses a dual-tower architecture built on a shared CodeT5 encoder. The test prefix and MUT are encoded separately with shared backbone weights and shared LoRA adapters. Mean pooling over non-padding tokens yields the test representation $h_t$ and the MUT representation $h_m$.

For \textsc{Pass}/\textsc{Fail} prediction, FOCAL constructs a relation-aware pair representation by concatenating the two representations, their absolute difference, and their element-wise product, denoted as $z=[h_t;h_m;|h_t-h_m|;h_t \odot h_m]$. The vector $z$ is passed to a lightweight classifier to produce the prediction logits and probabilities. This design preserves the separate roles of the test and the MUT while allowing the classifier to model their interaction.

For ranking supervision, our approach measures the compatibility between $h_t$ and $h_m$ using cosine similarity. The resulting score is used by the margin ranking loss to encourage compatible test--MUT pairs to receive higher scores than incompatible pairs.

\subsection{Fail-Aware Training}
As shown in Equation~\ref{eq:final-objective}, FOCAL combines fail-aware classification and margin ranking during training. The classification term is weighted by $\lambda_{\mathrm{cls}}$, and the ranking term is weighted by $\lambda_{\mathrm{rank}}$. Only the LoRA adapters and the lightweight classifier are updated, while the pretrained CodeT5 backbone remains frozen.

\begin{equation}
\mathcal{L}
=
\lambda_{\mathrm{cls}}\mathcal{L}_{\mathrm{cls}}
+
\lambda_{\mathrm{rank}}\mathcal{L}_{\mathrm{rank}}.
\label{eq:final-objective}
\end{equation}

\subsection{Statement-Level Evidence}

For pairs predicted as \textsc{Fail}, FOCAL identifies statement-level evidence in both the test and the MUT and evaluates whether the evidence is faithful to the model's decision. The explanation module consists of three steps: evidence extraction, faithfulness checks, and counterfactual-supported rationale presentation.\par

Given a test--MUT pair FOCAL first divides the test prefix and the MUT into statement-level units, assigns identifiers $T_i$ and $M_j$, and inserts explicit begin/end markers around each unit. Then it applies IG to the trained classifier with respect to the \textsc{Fail} logit. Attribution is computed separately for the test and MUT inputs, using a zero embedding baseline and fixed interpolation steps. Token-level scores are obtained by summing over embedding dimensions before being aggregated into statement-level scores according to tokenizer mappings and statement spans as depicted by the equation below:
\begin{equation}
A(S)
=
\sum_{r \in \mathcal{T}(S)}
w_{r,S} A_r ,
\label{eq:stmt-attr}
\end{equation}
where $\mathcal{T}(S)$ is the set of tokens overlapping statement $S$, $A_r$ is the token attribution score, and $w_{r,S}$ is the character-overlap weight. Statements are ranked by $A(S)$, and the top-$k$ statements are selected as the evidence set $E_k$, as shown in a concrete example in Figure~\ref{fig:running-example}, with \texttt{T3}, \texttt{M1}, \texttt{M4}.

\begin{figure*}[!t]
\centering
\begin{minipage}[t]{0.48\textwidth}
\textbf{Test prefix}
\begin{lstlisting}[language=Java,basicstyle=\ttfamily\scriptsize]
@Test
public void testBlankLine() throws Exception {
    StringBuilder sb = new StringBuilder();
    sb.append("/** Begin line maps. **/ { count : 1 }");
    sb.append(" ");                    // T3 evidence
    sb.append("/** Begin file information. **/ ");
    ...
}
\end{lstlisting}
\end{minipage}
\hfill
\begin{minipage}[t]{0.48\textwidth}
\textbf{Method under test}
\begin{lstlisting}[language=Java,basicstyle=\ttfamily\scriptsize]
void parse(String contents) throws SourceMapParseException {
    ParseState parser = new ParseState(contents); // M1 evidence
    try {
        parseInternal(parser);
    } catch (JSONException ex) {                  // M4 evidence
        parser.fail(...);
    }
}
\end{lstlisting}
\end{minipage}
\vspace{2pt}
\caption{Running example of FOCAL's statement-level evidence. FOCAL predicts \textsc{Fail} with $p_F=1.00$; Integrated Gradients ranks \texttt{M4}, \texttt{M1}, and \texttt{T3} as the top evidence; removing them lowers $p_F$ to $7.1\times10^{-3}$; keeping only them preserves $p_F=1.00$; the top-ranked counterfactual edit (deleting \texttt{M4}) flips the prediction to \textsc{Pass} with $p_F=0.06$.}
\label{fig:running-example}
\end{figure*}

FOCAL evaluates $E_k$ with two checks: a) removing the selected statements and b) keeping only the selected statements. The deletion check removes the selected evidence and measures the decrease in \textsc{Fail} confidence. As shown in Figure~\ref{fig:running-example}, after removing key evidence, the fail probability drops from 1.0 to $7.1 \times 10^{-3}$, a high drop indicating that the removed statements contribute more to the \textsc{Fail} prediction. And when keeping only key evidence statements, the probability can be kept at 1.0, a high keep ratio indicating that the selected statements preserve more of the evidence supporting the \textsc{Fail} prediction. 

For individual explanations, FOCAL optionally searches for a simple counterfactual edit over highly ranked statements. The search considers deleting a statement or replacing a MUT statement with a candidate statement from the original code. In the Figure~\ref{fig:running-example} example, deleting \texttt{M1} flips the prediction to \textsc{Pass} with $p_F = 0.06$. A successful edit changes the model prediction from \textsc{Fail} to \textsc{Pass}.

\section{Preliminary Evaluation}
\label{sec:evaluation}

We evaluate FOCAL with respect to prediction effectiveness and explanation usefulness through the following research questions:

\textbf{RQ1.} How accurately does FOCAL predict \textsc{Pass}/\textsc{Fail} outcomes and how does it performs with fail cases?

\textbf{RQ2.} To what extent do statement-level explanations truly reflect the evidence used in model failure predictions?

\subsection{Datasets}

We use a Defects4J-derived dataset~\cite{just2014defects4j} for in-domain evaluation. \textsc{Pass} instances pair a Defects4J test method with its original focal method, whereas \textsc{Fail} instances are constructed by applying mutation operators to the focal method, with filters removing invalid, trivial, and weak mutations. Across 17 projects and 854 bug versions, this pipeline yields 2.22M instances, which are split into training, validation, and test sets. The fail ratio does not reflect the frequency of failures in real executions; rather, it results from the mutation-based construction, which deliberately enriches failing cases to provide dense supervision.

For cross-project evaluation, we use an unseen-project benchmark constructed from TOGA RQ2 artifacts. \textsc{Pass} samples pair a focal method (C) with a test prefix (T). \textsc{Fail} samples are constructed from exception-oracle tests by deleting the \texttt{catch} block, causing the test to fail when the expected exception is not thrown. Table~\ref{tab:dataset-summary} reports the dataset scale, \textsc{Fail} ratio, and split policy. The in-domain split is group-aware, ensuring that instances from the same base test–MUT group are assigned to the same split. The unseen benchmark follows project-level separation from the training projects.

\begin{table}[!tb]
\caption{Dataset summary.}
\centering
\scriptsize
\setlength{\tabcolsep}{3pt}
\begin{tabular}{lllll}
\toprule
{Dataset} & {Proj.} & {Pairs} & {Fail \%} & {Split} \\
\midrule
D4J train    & 17 & 1{,}773{,}569 & 68.9 & Group-aware \\
D4J valid    & 17 & 222{,}657     & 68.9 & Group-aware \\
D4J test     & 17 & 222{,}278     & 68.9 & Group-aware \\
TOGA unseen  & 25 & 229{,}401     & 9.1  & Project-level \\
\bottomrule
\end{tabular}
\label{tab:dataset-summary}
\end{table}

Our approach was trained with LoRA rank 8, LoRA alpha 16, dropout 0.1, maximum length 512, batch size 8, learning rate $10^{-4}$, and trains for up to 10 epochs with early stopping. The training objective combines classification loss and margin ranking loss with $\lambda_{cls}=\lambda_{rank}=1.0$.\par

\subsection{RQ1: Prediction Effectiveness and Fail-Case Detection}

We treat \textsc{Pass} cases and \textsc{Fail} cases as two classes and report per-class Precision, Recall, and F1, along with overall Accuracy. FOCAL achieves strong in-domain performance on the Defects4J-derived validation and test sets, as shown in Table~\ref{tab:indomain-performance}. On the test set, it reaches 97.65\% accuracy, with 97.70\% Pass F1 and 97.61\% Fail F1. The close validation and test scores suggest that the model performs consistently within the same data construction setting.

\begin{table}[!tb]
\caption{In-domain performance of FOCAL.}
\centering
\scriptsize
\setlength{\tabcolsep}{3pt}
\begin{tabular}{llllllll}
\toprule
{Split} & {Pass Prec.} & {Pass Rec.} & {Pass F1} & {Fail Prec.} & {Fail Rec.} & {Fail F1} & {Acc.} \\
\midrule
Valid & 0.9583 & 0.9949 & 0.9763 & 0.9947 & 0.9567 & 0.9753 & 0.9758 \\
Test  & 0.9590 & 0.9956 & 0.9770 & 0.9955 & 0.9574 & 0.9761 & 0.9765 \\
\bottomrule
\end{tabular}
\label{tab:indomain-performance}
\end{table}

We further evaluate FOCAL on the unseen-project benchmark. On the pass-dominant unseen benchmark, SEER achieves higher overall accuracy than FOCAL, but this accuracy is largely driven by the majority PASS class. As shown in Table~\ref{tab:unseen-performance}, its FAIL recall is only 2.95\%, meaning that most failing cases are missed. In contrast, FOCAL sacrifices some aggregate accuracy but substantially improves failure detection: FAIL precision increases from 0.0565 to 0.1608, FAIL recall from 0.0295 to 0.2332, and FAIL F1 from 0.0387 to 0.1904. FOCAL also improves macro-F1 from 0.4837 to 0.5445, suggesting a more balanced behavior across PASS and FAIL cases. In absolute terms, FOCAL detects 4,856 failing cases, compared with approximately 614 detected by SEER.

\begin{table}[!tb]
\centering
\caption{Unseen-project performance comparison.}
\label{tab:unseen-performance}
\scriptsize
\setlength{\tabcolsep}{3pt}
\begin{tabular}{lcccccc}
\toprule
Model & Acc. & Pass F1 & Fail Prec. & Fail Rec. & Fail F1 & Macro F1 \\
\midrule
SEER & 0.8672 & 0.9287 & 0.0565 & 0.0295 & 0.0387 & 0.4837 \\
FOCAL & 0.8199 & 0.8987 & 0.1608 & 0.2332 & 0.1904 & 0.5445 \\
\bottomrule
\end{tabular}
\end{table}
Despite these improvements, FOCAL's fail-case detection remains limited in absolute terms. Its \textsc{Fail} recall and \textsc{Fail} F1 on unseen projects are still insufficient for dependable real-world use, where missed failures directly reduce the value of a learned oracle. We therefore view these results not as a solved problem, but as evidence of a promising and important research direction. Improving cross-project fail detection under realistic imbalanced settings remains a crucial challenge for future work.

\subsection{RQ2: Faithfulness of Statement-Level Explanations}

We evaluate explanations on 300 randomly sample pairs originally predicted as \textsc{Fail}. FOCAL ranks statements using signed IG and selects the top-$k$ statements as evidence. We compare this evidence with randomly selected statement sets of the same size.
For $k=3$, removing the top-ranked statements reduces \textsc{Fail} confidence by 0.3614 on average, compared with 0.0319 for random removal. Similarly, after removing the top-3 statements, the prediction flip rate is 52.33\%, and after removing the random statements, the flip rate is 6.00\%. In the keep-only check, top-ranked statements preserve most of the \textsc{Fail} confidence, with Keep Ratio close to or above 1.0 across $k \in \{1,3,5,10\}$.\par

The counterfactual check gives similar evidence. Searching over top-3 attributed statements flips 85.7\% of Fail predictions to \textsc{Pass}, compared with 10.3\% for random statements. Among successful top-3 edits, 178 flips come from deletion edits and 79 from replacement edits; the random baseline produces only 9 deletion flips and 22 replacement flips. These results suggest that the highlighted statements are not merely high-attribution artifacts, but are behaviorally connected to the model decision.

\section{Discussion}
\label{sec:discussion}

Our long-term vision goes beyond improving oracle prediction accuracy. We articulate this vision along three interconnected directions, including model-level improvement, integration with existing testing tools, and a broader paradigm shift in unit testing.

\textbf{Model level.} The cross-project results in this paper indicate that the most pressing improvement is generalization. A natural next step is to train and evaluate FOCAL on a more diverse mix of project families, generated and developer-written tests, real bugs, and mutation-based variants, and to study whether broader training diversity reduces this gap.\par
Another direction is to make explanations more actionable for developers. Future work can generate natural-language debugging hints or repair suggestions based on the identified key statements. Extend FOCAL toward developer-facing debugging support, making learned oracle predictions more usable in real-world unit testing.\par
\textbf{Integration with existing tools.} Another concrete instantiation of this vision is to integrate FOCAL with existing testing tools. Search-based tools, coverage-guided fuzzers, and LLM-based test generators can supply diverse executions or test prefixes; FOCAL might then act as a fault-detection layer over those candidates, flagging the subset whose behavior appears failure-prone and pointing to the statements most responsible for the model decision. The statement-level evidence FOCAL produces also fits naturally as input to downstream debugging techniques. For example, the highlighted statements can serve as starting points for delta debugging or spectrum-based fault localization, and the counterfactual edits used in our faithfulness checks can be viewed as model-level repair hypotheses rather than validated patches. We see these as concrete pipeline integrations, not as further claims of FOCAL itself. Furthermore, the same integration idea may also extend to other testing settings where executable inputs are available but reliable oracles remain difficult to obtain.\par

\textbf{Paradigm level.} Discriminative oracle prediction points a different way of using unit tests in AI-assisted testing workflows. Instead of treating a complete test case with an explicit oracle as the default generation target, future workflows may generate test prefixes that focus on inputs, boundary conditions, execution context, and domain scenarios, and then rely on learned oracle tools to support the oracle reasoning step: predicting \textsc{Pass}/\textsc{Fail} outcomes, explaining failures, highlighting suspicious statements, and eventually suggesting possible repairs. Such a shift reframes automatic test oracle generation from producing explicit oracle statements toward \emph{prefix test generation with AI-assisted oracle reasoning}. If realized, this paradigm might substantially reduce both human effort and server resources in unit testing.

\section{Limitations and Threats to Validity}

As an emerging result, FOCAL is subject to several threats to validity.\par
\textit{Construct validity.} A subset of \textsc{Fail} labels is constructed via mutation. Although mutants are a reasonable proxy for real defects~\cite{just2014mutants}, the gap is not fully closed.\par
\textit{Internal validity.} Because test prefixes and MUT variants may share common origins, we use group-aware splitting for in-domain evaluation and project-level separation for the unseen benchmark, to reduce leakage risk.
\textit{External validity.} The unseen benchmark is mainly pass-dominant, with only 9.1\% \textsc{Fail} cases; absolute \textsc{Fail} metrics should therefore be compared with care across datasets with different class ratios.\par
\textit{Explanation validity.} Our faithfulness checks establish behavioral grounding, but they do not demonstrate alignment with human reasoning; evaluation of extent of usefulness to developers remains future work.
Beyond these threats, improving the unseen performance through more diverse training corpora is an important next step.

\section{Conclusion}

This paper revisits discriminative test oracle prediction from a fail-aware perspective. Reviewing the state-of-the-art discriminative oracle construction work shows that high aggregate performance can still hide weak detection of failing cases, especially in unseen-project settings. FOCAL provides evidence from a preliminary evaluation that a fail-aware code LLM-based predictor can improve failure detection and produce statement-level evidence that is behaviorally checkable. The results are preliminary, but they point to a research direction worth exploring: learned oracles that directly reason about \textsc{Pass}/\textsc{Fail} behavior, especially where executable assertions are missing, unreliable, or costly to obtain.

\newpage
\bibliographystyle{ieeetr}
\newpage
\bibliography{references}

@book{myers2004art,
  title={The art of software testing},
  author={Myers, Glenford J and Badgett, Tom and Thomas, Todd M and Sandler, Corey},
  volume={2},
  year={2004},
  publisher={Wiley Online Library}
}

@book{ammann2017introduction,
  title={Introduction to software testing},
  author={Ammann, Paul and Offutt, Jeff},
  year={2017},
  publisher={Cambridge University Press}
}

@inproceedings{dinella2022toga,
  title={Toga: A neural method for test oracle generation},
  author={Dinella, Elizabeth and Ryan, Gabriel and Mytkowicz, Todd and Lahiri, Shuvendu K},
  booktitle={Proceedings of the 44th International Conference on Software Engineering},
  pages={2130--2141},
  year={2022}
}

@inproceedings{fraser2013evosuite,
  title={Evosuite: On the challenges of test case generation in the real world},
  author={Fraser, Gordon and Arcuri, Andrea},
  booktitle={2013 IEEE sixth international conference on software testing, verification and validation},
  pages={362--369},
  year={2013},
  organization={IEEE}
}

@inproceedings{pacheco2007randoop,
  title={Randoop: feedback-directed random testing for Java},
  author={Pacheco, Carlos and Ernst, Michael D},
  booktitle={Companion to the 22nd ACM SIGPLAN conference on Object-oriented programming systems and applications companion},
  pages={815--816},
  year={2007}
}

@article{barr2014oracle,
  title={The oracle problem in software testing: A survey},
  author={Barr, Earl T and Harman, Mark and McMinn, Phil and Shahbaz, Muzammil and Yoo, Shin},
  journal={IEEE transactions on software engineering},
  volume={41},
  number={5},
  pages={507--525},
  year={2014},
  publisher={IEEE}
}

@inproceedings{zhang2015assertions,
  title={Assertions are strongly correlated with test suite effectiveness},
  author={Zhang, Yucheng and Mesbah, Ali},
  booktitle={Proceedings of the 2015 10th Joint Meeting on Foundations of Software Engineering},
  pages={214--224},
  year={2015}
}

@book{sutton2007fuzzing,
  title={Fuzzing: brute force vulnerability discovery},
  author={Sutton, Michael and Greene, Adam and Amini, Pedram},
  year={2007},
  publisher={Pearson Education}
}

@inproceedings{shamshiri2015automatically,
  title={Do automatically generated unit tests find real faults? an empirical study of effectiveness and challenges (t)},
  author={Shamshiri, Sina and Just, Ren{\'e} and Rojas, Jos{\'e} Miguel and Fraser, Gordon and McMinn, Phil and Arcuri, Andrea},
  booktitle={2015 30th IEEE/ACM International Conference on Automated Software Engineering (ASE)},
  pages={201--211},
  year={2015},
  organization={IEEE}
}

@inproceedings{lemieux2023codamosa,
  title={Codamosa: Escaping coverage plateaus in test generation with pre-trained large language models},
  author={Lemieux, Caroline and Inala, Jeevana Priya and Lahiri, Shuvendu K and Sen, Siddhartha},
  booktitle={2023 IEEE/ACM 45th International Conference on Software Engineering (ICSE)},
  pages={919--931},
  year={2023},
  organization={IEEE}
}

@inproceedings{watson2020learning,
  title={On learning meaningful assert statements for unit test cases},
  author={Watson, Cody and Tufano, Michele and Moran, Kevin and Bavota, Gabriele and Poshyvanyk, Denys},
  booktitle={Proceedings of the ACM/IEEE 42nd International Conference on Software Engineering},
  pages={1398--1409},
  year={2020}
}

@inproceedings{yu2022automated,
  title={Automated assertion generation via information retrieval and its integration with deep learning},
  author={Yu, Hao and Lou, Yiling and Sun, Ke and Ran, Dezhi and Xie, Tao and Hao, Dan and Li, Ying and Li, Ge and Wang, Qianxiang},
  booktitle={Proceedings of the 44th International Conference on Software Engineering},
  pages={163--174},
  year={2022}
}

@inproceedings{blasi2018translating,
  title={Translating code comments to procedure specifications},
  author={Blasi, Arianna and Goffi, Alberto and Kuznetsov, Konstantin and Gorla, Alessandra and Ernst, Michael D and Pezz{\`e}, Mauro and Castellanos, Sergio Delgado},
  booktitle={Proceedings of the 27th ACM SIGSOFT international symposium on software testing and analysis},
  pages={242--253},
  year={2018}
}

@inproceedings{ibrahimzada2022perfect,
  title={Perfect is the enemy of test oracle},
  author={Ibrahimzada, Ali Reza and Varli, Yigit and Tekinoglu, Dilara and Jabbarvand, Reyhaneh},
  booktitle={Proceedings of the 30th acm joint european software engineering conference and symposium on the foundations of software engineering},
  pages={70--81},
  year={2022}
}

@inproceedings{wang2021codet5,
  title={Codet5: Identifier-aware unified pre-trained encoder-decoder models for code understanding and generation},
  author={Wang, Yue and Wang, Weishi and Joty, Shafiq and Hoi, Steven CH},
  booktitle={Proceedings of the 2021 conference on empirical methods in natural language processing},
  pages={8696--8708},
  year={2021}
}

@inproceedings{sundararajan2017axiomatic,
  title={Axiomatic attribution for deep networks},
  author={Sundararajan, Mukund and Taly, Ankur and Yan, Qiqi},
  booktitle={International conference on machine learning},
  pages={3319--3328},
  year={2017},
  organization={PMLR}
}

@inproceedings{motwani2023better,
  title={Better automatic program repair by using bug reports and tests together},
  author={Motwani, Manish and Brun, Yuriy},
  booktitle={2023 IEEE/ACM 45th International Conference on Software Engineering (ICSE)},
  pages={1225--1237},
  year={2023},
  organization={IEEE}
}

@inproceedings{nilizadeh2021exploring,
  title={Exploring true test overfitting in dynamic automated program repair using formal methods},
  author={Nilizadeh, Amirfarhad and Leavens, Gary T and Le, Xuan-Bach D and P{\u{a}}s{\u{a}}reanu, Corina S and Cok, David R},
  booktitle={2021 14th IEEE conference on software testing, verification and validation (ICST)},
  pages={229--240},
  year={2021},
  organization={IEEE}
}

@inproceedings{almasi2017industrial,
  title={An industrial evaluation of unit test generation: Finding real faults in a financial application},
  author={Almasi, M Moein and Hemmati, Hadi and Fraser, Gordon and Arcuri, Andrea and Benefelds, Janis},
  booktitle={2017 IEEE/ACM 39th International Conference on Software Engineering: Software Engineering in Practice Track (ICSE-SEIP)},
  pages={263--272},
  year={2017},
  organization={IEEE}
}

@inproceedings{fioraldi2020afl++,
  title={$\{$AFL++$\}$: Combining incremental steps of fuzzing research},
  author={Fioraldi, Andrea and Maier, Dominik and Ei{\ss}feldt, Heiko and Heuse, Marc},
  booktitle={14th USENIX workshop on offensive technologies (WOOT 20)},
  year={2020}
}

@inproceedings{serebryany2016continuous,
  title={Continuous fuzzing with libfuzzer and addresssanitizer},
  author={Serebryany, Kosta},
  booktitle={2016 IEEE Cybersecurity Development (SecDev)},
  pages={157--157},
  year={2016},
  organization={IEEE}
}

@inproceedings{padhye2019jqf,
  title={Jqf: Coverage-guided property-based testing in java},
  author={Padhye, Rohan and Lemieux, Caroline and Sen, Koushik},
  booktitle={Proceedings of the 28th ACM SIGSOFT International Symposium on Software Testing and Analysis},
  pages={398--401},
  year={2019}
}

@inproceedings{tufano2022generating,
  title={Generating accurate assert statements for unit test cases using pretrained transformers},
  author={Tufano, Michele and Drain, Dawn and Svyatkovskiy, Alexey and Sundaresan, Neel},
  booktitle={Proceedings of the 3rd ACM/IEEE International Conference on Automation of Software Test},
  pages={54--64},
  year={2022}
}

@article{alagarsamy2024a3test,
  title={A3test: Assertion-augmented automated test case generation},
  author={Alagarsamy, Saranya and Tantithamthavorn, Chakkrit and Aleti, Aldeida},
  journal={Information and Software Technology},
  volume={176},
  pages={107565},
  year={2024},
  publisher={Elsevier}
}

@inproceedings{hossain2025togll,
  title={Togll: Correct and strong test oracle generation with llms},
  author={Hossain, Soneya Binta and Dwyer, Matthew B},
  booktitle={2025 IEEE/ACM 47th International Conference on Software Engineering (ICSE)},
  pages={1475--1487},
  year={2025},
  organization={IEEE}
}

@article{schafer2023empirical,
  title={An empirical evaluation of using large language models for automated unit test generation},
  author={Sch{\"a}fer, Max and Nadi, Sarah and Eghbali, Aryaz and Tip, Frank},
  journal={IEEE Transactions on Software Engineering},
  volume={50},
  number={1},
  pages={85--105},
  year={2023},
  publisher={IEEE}
}

@inproceedings{siddiq2024using,
  title={Using large language models to generate junit tests: An empirical study},
  author={Siddiq, Mohammed Latif and Da Silva Santos, Joanna Cecilia and Tanvir, Ridwanul Hasan and Ulfat, Noshin and Al Rifat, Fahmid and Carvalho Lopes, Vin{\'\i}cius},
  booktitle={Proceedings of the 28th international conference on evaluation and assessment in software engineering},
  pages={313--322},
  year={2024}
}

@inproceedings{lin2017focal,
  title={Focal loss for dense object detection},
  author={Lin, Tsung-Yi and Goyal, Priya and Girshick, Ross and He, Kaiming and Doll{\'a}r, Piotr},
  booktitle={Proceedings of the IEEE international conference on computer vision},
  pages={2980--2988},
  year={2017}
}

@article{wachter2017counterfactual,
  title={Counterfactual explanations without opening the black box: Automated decisions and the GDPR},
  author={Wachter, Sandra and Mittelstadt, Brent and Russell, Chris},
  journal={Harv. JL \& Tech.},
  volume={31},
  pages={841},
  year={2017},
  publisher={HeinOnline}
}

@inproceedings{just2014defects4j,
  title={Defects4J: A database of existing faults to enable controlled testing studies for Java programs},
  author={Just, Ren{\'e} and Jalali, Darioush and Ernst, Michael D},
  booktitle={Proceedings of the 2014 international symposium on software testing and analysis},
  pages={437--440},
  year={2014}
}

@inproceedings{just2014mutants,
  title={Are mutants a valid substitute for real faults in software testing?},
  author={Just, Ren{\'e} and Jalali, Darioush and Inozemtseva, Laura and Ernst, Michael D and Holmes, Reid and Fraser, Gordon},
  booktitle={Proceedings of the 22nd ACM SIGSOFT international symposium on foundations of software engineering},
  pages={654--665},
  year={2014}
}
\end{document}